\documentclass[10pt]{amsart}

\usepackage{geometry}
\usepackage{parskip} % NOT "\setlength{\parindent}{0pt}"

\usepackage{graphicx}
\usepackage{amssymb}
\usepackage{epstopdf}
\usepackage{url}
\usepackage{subfigure}
\usepackage{lineno}
\usepackage{color}

\usepackage{placeins}
\usepackage{amsmath}
\usepackage{mathtools}
\usepackage{color, colortbl}
\usepackage{fancyhdr}
\usepackage{ctable}
\definecolor{Gray}{gray}{0.9}
\usepackage{enumitem}
\usepackage{graphicx}
\usepackage{caption}
\usepackage{color, colortbl}
\usepackage{array}
\usepackage{booktabs} % For formal tables
\usepackage{makecell}
\usepackage{cellspace}
\usepackage{tabularx}
\usepackage{multirow}
\usepackage{natbib}
\usepackage[foot]{amsaddr}
\usepackage{endnotes}

\let\footnote=\endnote

\begin{document}

\title[How Epidemic Psychology Works on Twitter]{How Epidemic Psychology Works on Twitter:\\Evolution of responses to the COVID-19 pandemic in the U.S.}

\author{Luca Maria Aiello$^{1,2}$}
\author{Daniele Quercia$^{1,3,*}$}
\author{Ke Zhou$^{1}$}
\author{Marios Constantinides$^{1}$}
\author{Sanja Scepanovic$^{1}$}
\author{Sagar Joglekar$^{1}$}

\address{$^{1}$Nokia Bell Labs Cambridge, United Kingdom}
\address{$^{2}$IT University Copenhagen, Denmark}
\address{$^{3}$Centre for Urban Science and Progress, King's College London, United Kingdom}
\address{$^{*}$Corresponding author: quercia@cantab.net}

\begin{abstract}
Disruptions resulting from an epidemic might often appear to amount to chaos but, in reality, can be understood in a systematic way through the lens of ``epidemic psychology''. According to Philip Strong, the founder of the sociological study of epidemic infectious diseases, not only is an epidemic biological; there is also the potential for three psycho-social epidemics: of fear, moralization, and action. This work empirically tests Strong's model at scale by studying the use of language of 122M tweets related to the COVID-19 pandemic posted in the U.S. during the whole year of 2020. On Twitter, we identified three distinct phases. Each of them is characterized by different regimes of the three psycho-social epidemics. In the refusal phase, users refused to accept reality despite the increasing number of deaths in other countries. In the anger phase (started after the announcement of the first death in the country), users' fear translated into anger about the looming feeling that things were about to change. Finally, in the acceptance phase, which began after the authorities imposed physical-distancing measures, users settled into a ``new normal'' for their daily activities. Overall, refusal of accepting reality gradually died off as the year went on, while acceptance increasingly took hold. During 2020, as cases surged in waves, so did anger, re-emerging cyclically at each wave. Our real-time operationalization of Strong's model is designed in a way that makes it possible to embed epidemic psychology into real-time models (e.g., epidemiological and mobility models). 
\end{abstract}

\flushbottom
\maketitle
\thispagestyle{empty}

%%%%%%%%%%%%%%%%%%%%%%%%%%%%%%%%%%%%%%%
%%%%%%%%%%%%%%%%%%%%%%%%%%%%%%%%%%%%%%%
\section*{Introduction}
%%%%%%%%%%%%%%%%%%%%%%%%%%%%%%%%%%%%%%%
%%%%%%%%%%%%%%%%%%%%%%%%%%%%%%%%%%%%%%%

In our daily lives, our dominant perception is of order. But every now and then chaos threatens that order: epidemics dramatically break out, revolutions erupt, empires suddenly fall, and stock markets crash. Epidemics, in particular, present not only collective health hazards but also special challenges to mental health and public order that need to be addressed by social and behavioral sciences~\citep{van2020using}. Almost 30 years ago, in the wake of the AIDS epidemic, Philip Strong, the founder of the sociological study of epidemic infectious diseases, reflected: \emph{``the human origin of epidemic psychology lies not so much in our unruly passions as in the threat of epidemic disease to our everyday assumptions.''}~\citep{strong1990epidemic} In the recent COVID-19 pandemic~\citep{brooks2020psychological} (an ongoing pandemic of a coronavirus disease), it has been shown that the main source of uncertainty and anxiety has indeed come from the disruption of what Alfred Shutz called the ``routines and recipes'' of daily life~\citep{schutz1973structures} (e.g., every simple act, from eating at work to visiting our parents, takes on new meanings).

Yet, the chaos resulting from an epidemic turns out to be more predictable than what one would initially expect. Philip Strong observed that any new \emph{health} epidemic resulted into three \emph{psycho-social} epidemics: of fear, moralization, and action. The epidemic of fear represents the fear of catching the disease, which comes with the suspicion against alleged disease carriers, which, in turn, may spark panic and irrational behavior. The epidemic of moralization is characterized by moral responses both to the viral epidemic itself and to the epidemic of fear, which may result in either positive reactions (e.g., cooperation) or negative ones (e.g., stigmatization). The epidemic of action accounts for the rational or irrational changes of daily habits that people make in response to the disease or as a result of the two other psycho-social epidemics. Strong was writing in the wake of the AIDS/HIV crisis, but he based his model on studies that went back to Europe's Black Death in the $14^{th}$ century. Importantly, he showed that these three psycho-social epidemics are created by language and incrementally fed through it: language transmits the \emph{fear} that the infection is an existential threat to humanity and that we are all going to die; language depicts the epidemic as a verdict on human failings and as a \emph{moral} judgment on minorities; and language shapes the means through which people collectively intend to, however pointless, \emph{act} against the threat. 

There have been numerous studies of how information propagated on social media during epidemic outbreaks occurred in the past decade, such as Zika~\citep{fu2016people,wood2018propagating,sommariva2018spreading}, Ebola~\citep{oyeyemi2014ebola}, and the H1N1 influenza~\citep{chew2010pandemics}. Similarly, with the outbreak of COVID-19, people around the world have been collectively expressing their thoughts and concerns about the pandemic on social media. As such, researchers studied this epidemic from multiple angles: social media posts have been analyzed in terms of content and behavioral markers~\citep{hou2020assessment,li2020impact}, and of tracking the diffusion of COVID-related information~\citep{cinelli2020covid} and misinformation~\citep{pulido2020covid,ferrara2020covid,kouzy2020coronavirus,yang2020prevalence}. Search queries have suggested specific information-seeking responses to the pandemic~\citep{bento2020evidence}. Psychological responses to COVID-19 have been studied mostly though surveys~\citep{wang2020immediate,qiu2020nationwide}.

Hitherto, there has never been any large-scale empirical study of whether the use of language during an epidemic reflects Strong's model. With the opportunity of having a sufficiently large-scale data at hand, we set out to test whether Strong's model did hold on Twitter during COVID-19, and did so at the scale of an entire country, that of the United States. By running our study on Twitter, we  expose the interpretation of our results to a number of limitations, notably to issues of representativeness~\citep{li2013spatial}, self-presentation biases~\citep{waterloo2018norms}, and data noise~\citep{saif2012alleviating,ferrara2016rise}. Indeed, recent surveys estimated that only 22\% of U.S. adults use Twitter~\citep{wojcik2019sizing}, and that the characteristics of these users deviate from the general population's:  compared to the average adult in the U.S., Twitter users are much younger, are more likely to have college degrees, and are slightly more likely to identify with the Democratic Party. Despite such limitations, social media represents the response of a relevant part of the general population  to global events, and it does so at a scale and a granularity that have been so far unattainable by publicly-available data sources.
	
After operationalizing Strong's model by using lexicons for psycholinguistic text analysis, upon the collection of 122M tweets about the epidemic from February $1^{st}$ to December $31^{st}$, we conducted a quantitative analysis on the differences in language style and a thematic analysis of the actual social media posts. The temporal scope of our study does not capture the full pandemic as it  is still ongoing at the time of writing but it includes the three major contagion waves in the U.S., characterizing the entire year of 2020. The first wave captured the initial diffusion of the virus in the world, and then its arrival and first peak in the U.S. The following two waves captured subsequent periods of alarming diffusion. 

The three psycho-social epidemics, as theorized by Strong, evolve concurrently over time. In the particular case of our period of study, we found that this concurrent evolution resulted into three regimes or phases, which are not part of Strong's theoretical framework and experimentally emerged. In the first phase (the \emph{refusal} phase), the psycho-social epidemic of fear began. Twitter users in the U.S. refused to accept reality: they feared the uncertainty created by the disruption of what was considered to be ``normal''; focused their moral concerns on others in an act of distancing oneself from others; yet, despite all this, they refused to change the normal course of action. After the announcement of the first death in the country, the second phase (the \emph{anger} phase) began: the psycho-social epidemic of fear intensified while the epidemics of morality and action kicked-off abruptly. Twitter users expressed more anger than fear about the looming feeling that things were about to change; focused their moral concerns on oneself in an act of reckoning with what was happening; and suspended their daily activities. After the authorities imposed physical-distancing measures, the third phase (the \emph{acceptance} phase) took over: the epidemic of fear started to fade away while the epidemics of morality and action turned into more constructive and forward-looking social processes. Twitter users expressed more sadness than anger or fear; focused their moral concerns on the collective and, in so doing, promoted pro-social behavior; and found a ``new normal'' in their  daily activities, which consisted of their past daily activities being  physically restricted to their homes and neighborhoods. The phase of \emph{acceptance} dominated Twitter conversations for the rest of the year, although the \emph{anger} phase re-emerged cyclically with the rise of new waves of contagion. In particular, we observed two peaks of anger: when the death toll in the U.S. reached 100,000 people (second contagion wave), and when President Trump tested positive to COVID-19 (third contagion wave).

\section*{Dataset} \label{sec:data-collection}
From an existing collection of COVID-related tweets~\citep{chen2020covid}, we gathered 554,941,519 tweets posted between February $1^{st}$ up to December $31^{st}$. We focused our analysis on the United States, the country where Twitter penetration is highest. To identify Twitter users living in it, we parsed the free-text location description of their user profile (e.g.,~``San Francisco, CA''). We did so by using a set of custom regular expressions that match variations for the expression ``United States of America'', as well as the names of 333 cities and 51 states in the U.S. (and their combinations). Albeit not always accurate, matching location strings against known location names is a tested approach that yields good results for a coarse-grained localization at state or country-level~\citep{dredze2013carmen}. Overall, we were left with 6,271,835 unique users in  the U.S. who posted 122,320,155 tweets in English.

Before analyzing  how our language categories unfolded over time, we experimentally tested 
whether the number of data points at hand was sufficient to compute our metrics. It was indeed  the case as the number of active users per day varied from a minimum of 72k on February $2^{nd}$ to a maximum of 1.84M on March $18^{th}$, with and average of 437k. A small number of accounts tweeted a disproportionately high number of times, reaching a maximum of 15,823 tweets; those were clearly automated accounts, which were discarded by our methodology. As we shall discuss in Methods, we normalized all our aggregate temporal measures so that they were not affected by the fluctuating volume of tweets over time.

%%%%%%%%%%%%%%%%%%%%%%%%%%%%%%%%%%%%%%%
%%%%%%%%%%%%%%%%%%%%%%%%%%%%%%%%%%%%%%%
\section*{Methods} \label{sec:methods}
%%%%%%%%%%%%%%%%%%%%%%%%%%%%%%%%%%%%%%%
%%%%%%%%%%%%%%%%%%%%%%%%%%%%%%%%%%%%%%%

\subsection*{Coding Strong's model} \label{sec:methods:strong}

Back in the 1990s, Philip Strong was able not only to describe the psychological impact of epidemics on social order but also to model it. He observed that the early reaction to major fatal epidemics is a distinctive psycho-social form and can be modeled along three main dimensions: fear, morality, and action. During a large-scale epidemic, basic assumptions about social interaction and, more generally, about social order are disrupted, and, more specifically, they are so by: the fear of others, competing moralities, and the responses to the epidemic. Crucially, all these three elements are created, transmitted, and mediated by language: language transmits fears, elaborates on the stigmatization of minorities, and shapes the means through which people collectively respond to the epidemic~\citep{strong1990epidemic,goffman2009stigma,cap2016language}.

As opposed to existing attempts to model psychological and social aspects of epidemic crises~\citep{mcconnell2005banks,khan2019psychology}, Strong's model meets our three main choice criteria:
\begin{enumerate}
	\item \emph{Well-grounded}. It proposes  a comprehensive, highly-cited, and still relevant theoretical model, which is based on an extensive review of studies of past large-scale epidemics 
that did span centuries and, as such, were of different nature, speaking to the generalizability of the framework.
	\item \emph{Focused on psycho-social aspects}. Its main goal is to characterize people's psychological and social responses to epidemics rather than describing how an epidemic unfolds over time.
	\item \emph{Directly operationalizable from language use}. The description of the psycho-social responses provided by Strong lends itself to operationalization, as its defining concepts  have been mapped to language markers by previous literature (as Table~\ref{table:literature_review} shows). 
\end{enumerate}

We operationalized Strong's epidemic psychology theoretical framework in two steps. First, three authors hand-coded Strong's seminal paper~\citep{strong1990epidemic} using \emph{line-by-line} coding~\citep{gibbs2007thematic} to identify keywords that characterize the three psycho-social epidemics. For each of the three psycho-social epidemics, the three authors generated independent lists of keywords that were conservatively combined by intersecting them. The words that were left out by the intersection were mostly synonyms (e.g., ``catching disease'' as a synonym for ``contagion''), so we did not discard any important concept. According to Strong, the three psycho-social epidemics are intertwined and, as such, the concepts that define one specific psycho-social epidemic might be relevant to the remaining two as well. For example, \emph{suspicion} is an element of the epidemic of fear but is tightly related to \emph{stigmatization} as well, a phenomenon that Strong describes as typical of the epidemic of moralization. In our coding exercise, we adhered as much as possible to the description in Strong's paper and obtained a strict partition of keywords across psycho-social epidemics. In the second step, the same three authors mapped each of these keywords to \emph{language categories}, namely sets of words that reflect how these concepts are expressed in natural language (e.g., words expressing \emph{anger} or \emph{trust}). We took these categories from four existing language lexicons widely used in psychometric studies: 
\begin{itemize}
	\item \emph{Linguistic Inquiry Word Count (LIWC)~\citep{tausczik2010psychological}}. A lexicon of words and word stems grouped into over 125 categories reflecting emotions, social processes, and basic functions, among others. The LIWC lexicon is based on the premise that the words people use to communicate can provide clues to their psychological states~\citep{tausczik2010psychological}. It allows written passages to be analyzed syntactically (how the words are used together to form phrases or sentences) and semantically (an analysis of the meaning of the words or phrases).
	\item \emph{Emolex~\citep{mohammad2013crowdsourcing}}. A lexicon that classifies 6k+ words and stems into the eight primary emotions of Plutchik's psychoevolutionary theory~\citep{plutchik1991emotions}.
	\item \emph{Moral Foundation Lexicon~\citep{graham2009liberals}}. A lexicon of 318 words and stems, which are grouped into 5 categories of moral foundations~\citep{graham2013moral}: harm, fairness, in-group, authority, and purity. Each of which is further split into expressions of virtue or vice.
	\item \emph{Pro-social behavior~\citep{frimer2014moral}}. A lexicon of 146 pro-social words and stems, which have been found to be frequently used when people describe pro-social goals\citep{frimer2014moral}.
\end{itemize}
The three authors grouped similar keywords together and mapped groups of keywords to one or more language categories. This grouping and mapping procedure was informed by previous studies that investigated how these keywords are expressed through language. These studies are listed in Table~\ref{table:literature_review}.

{\def\arraystretch{1.5}
\begin{table*}[ht!]
\setlength{\tabcolsep}{1.5mm}
\centering
\scriptsize
\begin{tabular}{c p{30mm}  p{70mm} p{27mm} p{6mm} p{6mm} p{6mm}}
\specialrule{.1em}{.05em}{.05em} 

 & & & & \multicolumn{3}{c}{\textbf{Max. peak}} \\
 & \multicolumn{1}{c}{\textbf{Keywords}} & \multicolumn{1}{c}{\textbf{Supporting literature}} & \textbf{Lang. categories}  & \multicolumn{1}{c}{\textbf{1$^{st}$}} & \multicolumn{1}{c}{\textbf{2$^{nd}$}}  & \multicolumn{1}{c}{\textbf{3$^{rd}$}} \\
\Xhline{2\arrayrulewidth}
%%%%%%%%%%%%%%%%%%%%%%%%%%%
\multirow{16}{*}{\rotatebox[origin=c]{90}{\textbf{Fear}}} 

		 & \cellcolor{Gray} emotional maelstrom & \cellcolor{Gray} & \cellcolor{Gray}swear (liwc)   & \cellcolor{Gray}.03 & \cellcolor{Gray} \textbf{.54} & \cellcolor{Gray}.14 \\ 
		& \cellcolor{Gray} & \cellcolor{Gray}  & \cellcolor{Gray} anger (liwc)   & \cellcolor{Gray} .07 & \cellcolor{Gray} .14 & \cellcolor{Gray} \textbf{.16} \\		
		& \cellcolor{Gray} &  \cellcolor{Gray} & \cellcolor{Gray} negemo (liwc)  & \cellcolor{Gray} .09 & \cellcolor{Gray} \textbf{.17} & \cellcolor{Gray} .02 \\
		& \cellcolor{Gray} & \multirow{-4}{=}{\cellcolor{Gray}\setlength\parskip{\baselineskip}These LIWC categories have been used to analyze complex emotional responses to traumatic events (PTSD) and to characterize the language of people suffering from mental health~\citep{coppersmith2014quantifying}\endgraf}  & \cellcolor{Gray} sadness (liwc) & \cellcolor{Gray} -.09 & \cellcolor{Gray} .04 & \cellcolor{Gray} \textbf{.19} \\
		%anxiety
		
     & fear & \multirow{2}{=}{\setlength\parskip{\baselineskip}Fear-related words like the ones included in Emolex have been often used to measure fear of both tangible and intangible threats~\citep{kahn2007measuring,gill2008language}\endgraf}
		 & fear (emolex) & \textbf{.07} & .05 & .03 \\
		 & & & death (liwc) & \textbf{.45} & .16 & .03 \\
		
		 & \cellcolor{Gray} anxiety, panic & \cellcolor{Gray} The \emph{anxiety} category of LIWC has been used to study different forms of anxiety in social media~\citep{shen2017detecting} & \cellcolor{Gray} anxiety (liwc) & \cellcolor{Gray} .31 & \cellcolor{Gray} \textbf{.46} & \cellcolor{Gray} -.15  \\
		
     & disorientation & By definition, the \emph{tentative} category of LIWC expresses uncertainty~\citep{tausczik2010psychological} & tentative (liwc) & .02 & \textbf{.10} & .03 \\
		
		 & \cellcolor{Gray} suspicion & \cellcolor{Gray} Suspicion is often formalized as lack of trust~\citep{deutsch1958trust} & \cellcolor{Gray} trust (liwc)  & \cellcolor{Gray} -.04 &  \cellcolor{Gray} .03 & \cellcolor{Gray} \textbf{.12}\\
		
		 & irrationality & The \emph{negate} category of LIWC has been used to measure cognitive distorsions and irrational interpretations of reality~\citep{simms2017detecting} & negate (liwc) & .08 & \textbf{.15} & .07 \\
		
		& \cellcolor{Gray} religion & \cellcolor{Gray} Religious expressions from LIWC have been used to study how people appeal to religious entities during moments of hardship~\citep{shaw2007effects} & \cellcolor{Gray} religion (liwc)  & \cellcolor{Gray} .12 & \cellcolor{Gray} .17 & \cellcolor{Gray} \textbf{.22}\\
		
		& contagion & \multirow{3}{=}{\setlength\parskip{\baselineskip}These LIWC categories were used to study the perception of diseases in cancer support groups, people affected by eating disorder, and alcoholics~\citep{alpers2005evaluation,wolf2013language,kornfield2018you}\endgraf}
			   & body (liwc)  & .01 & \textbf{.27} & .13 \\
			& & & feel (liwc)  & -.04 & \textbf{.34} & .03 \\
			& & &   &  & &  \\

%%%%%%%%%%%%%%%%%%%%%%%%%%%
\specialrule{.1em}{.05em}{.05em} 
\multirow{12}{*}{\rotatebox[origin=c]{90}{\textbf{Moralization}}}

	& \cellcolor{Gray} warn, risk avoidance, risk perception & \cellcolor{Gray} A LIWC category used to model risk perception connected to epidemics~\citep{hou2020assessment} & \cellcolor{Gray}risk (liwc)  & \cellcolor{Gray} .04 & \cellcolor{Gray} \textbf{.15} & \cellcolor{Gray} .08 \\
	
							& polarization, segregation & \multirow{4}{=}{\setlength\parskip{\baselineskip}Different personal pronouns have been used to study in- and out-group dynamics and to characterize language markers of racism~\citep{arguello2006talk}; personal pronouns and markers of differentiation have been considered in studies on racist language~\citep{figea2016measuring}\endgraf}
						    & I (liwc)      & -.10 & \textbf{.49} & .26 \\
						& & & we (liwc)     & -.09 & \textbf{.24} & .22 \\
						& & & they (liwc)   & .03 & .02 & \textbf{.10} \\
						& & & differ (liwc) &  .01 & \textbf{.08} & .05 \\
	
						 & \multirow{1}{=}{\cellcolor{Gray}\setlength\parskip{\baselineskip}stigmatization, blame, abuse\endgraf} &  \cellcolor{Gray} Pronouns \emph{I} and \emph{they} were used to quantify blame in personal~\citep{borelli2011trauma} and political context~\citep{windsor2014language}. Hate speech is associated with they-oriented statements~\citep{elsherief2018hate} 
						    & \cellcolor{Gray} (same categories as previous line) & \cellcolor{Gray} & \cellcolor{Gray} & \cellcolor{Gray}\\
						
						& \multirow{3}{=}{\setlength\parskip{\baselineskip}cooperation, coordination, collective consciousness\endgraf} & \multirow{3}{=}{\setlength\parskip{\baselineskip}The moral value of \emph{care} expresses the will of protecting others~\citep{graham2009liberals}. Cooperation is often verbalized by referencing the in-group and by expressing \emph{affiliation}~\citep{rezapour2019enhancing}\endgraf}
						&  affiliation (liwc) & -.16 & \textbf{.25} & .22 \\
						&  &  & care (moral virtue) & -.09  & .12 &  \textbf{.19} \\
						&  &  &  prosocial (prosocial) & -.10 & .07  & \textbf{.25} \\
						
						& \cellcolor{Gray}faith in authority & \cellcolor{Gray}The moral value of \emph{authority} expresses the will of playing by the rules of a hierarchy versus challenging it~\citep{rezapour2019enhancing} & \cellcolor{Gray}authority (moral virtue)  & \cellcolor{Gray} -.04 & \cellcolor{Gray} .11 & \cellcolor{Gray} \textbf{.13} \\
						
						& authority enforcement & The \emph{power} category of LIWC expresses exertion of dominance~\citep{tausczik2010psychological} & power (liwc)   & -.01 & .02 & \textbf{.14}\\
				
\specialrule{.1em}{.05em}{.05em} 
%%%%%%%%%%%%%%%%%%%%%%%%%%%
\multirow{5}{*}{\rotatebox[origin=c]{90}{\textbf{Action}}}     
          & \cellcolor{Gray} restrictions, travel, privacy & \cellcolor{Gray} & \cellcolor{Gray}motion (liwc) & \cellcolor{Gray} .02 & \cellcolor{Gray} .04 & \cellcolor{Gray} \textbf{.06}\\
					& \cellcolor{Gray} & \cellcolor{Gray} & \cellcolor{Gray}home (liwc) & \cellcolor{Gray} -.15 & \cellcolor{Gray} \textbf{.38} & \cellcolor{Gray} .20 \\
					& \cellcolor{Gray} &\cellcolor{Gray} & \cellcolor{Gray} work (liwc) & \cellcolor{Gray} -.05 & \cellcolor{Gray} .04 & \cellcolor{Gray} \textbf{.22} \\
					& \cellcolor{Gray} & \cellcolor{Gray} & \cellcolor{Gray} social (liwc) &\cellcolor{Gray} -.06  & \cellcolor{Gray} \textbf{.09} & \cellcolor{Gray} .08\\
					& \cellcolor{Gray} & \multirow{-8}{=}{\cellcolor{Gray}\setlength\parskip{\baselineskip}Daily habits concern mainly people's experience of \emph{home}, \emph{work}, \emph{leisure}, and \emph{movement} between them~\citep{gonzalez2008understanding}\endgraf} & \cellcolor{Gray} leisure (liwc)   & \cellcolor{Gray} .05 & \cellcolor{Gray} \textbf{.16} & \cellcolor{Gray} .10\\
%\Xhline{2\arrayrulewidth}
\specialrule{.1em}{.05em}{.05em}

\end{tabular}
\captionsetup{width=.99\linewidth}
\caption{Operationalization of the Strong's epidemic psychology theoretical framework. From Strong's paper, three annotators extracted \emph{keywords} that characterize the three psycho-social epidemics and mapped them to relevant \emph{language categories} from existing language lexicons used in psychometric studies. Category names are followed by the name of their corresponding lexicon in parenthesis. We support the association between keywords and language categories with examples of \emph{supporting literature}. To summarize how the use of the language categories varies across the three temporal states, we computed the \emph{peak} values of the different language categories (days when their standardized fractions reached the maximum), and reported the percentage increase at peak compared to the average over the whole time period; in each row, the maximum value is highlighted in bold.}
\label{table:literature_review}
\vspace{35pt}
\end{table*}
}

\subsection*{Language categories over time} \label{sec:quantitative-analysis}

We considered that a tweet contained a language category $c$ if at least one of the tweet's words or stems belonged to that category. The tweet-category association is binary and disregards the number of matching words within the same tweet. That is mainly because, in short snippets of text (tweets are limited to 280 characters), multiple occurrences are rare and do not necessarily reflect the intensity of a category~\citep{russell2013mining}. For each language category $c$, we counted the number of users $U_c(t)$ who posted at least one tweet at time $t$ containing that category. We then obtained the fraction of users who mentioned category $c$ by dividing $U_c(t)$ by the total number of users $U(t)$ who tweeted at time $t$:
\begin{equation}
f_c(t) = \frac{U_c(t)}{U(t)}.
\label{eqn:f_d_t}
\end{equation}
Computing the fraction of users rather than the fraction of tweets prevents biases introduced by exceptionally active users, thus capturing more faithfully the prevalence of different language categories in our Twitter population. This also helps discounting the impact of social bots, which tend to have anomalous levels of activity (especially retweeting~\citep{bessi2016social}).

Different categories might be verbalized with considerably different frequencies. For example, the language category ``I'' (first-person pronoun) from the LIWC lexicon naturally occurred much more frequently than the category ``death'' from the same lexicon. To enable a comparison across categories, we standardized all the fractions:
\begin{equation}
z_c(t) =  \frac{f_c(t) - \mu_{[0,T]}(f_c)}{\sigma_{[0,T]}(f_c)},
\label{eqn:z_d_t}
\end{equation}
where $\mu(f_c)$ and $\sigma(f_c)$ represent the mean and standard deviation of the $f_c(t)$ scores over the whole time period, from $t=0$ (February $1^{st}$) to $t=T$ (April $16^{th}$). These $z$-scores ease also the interpretation of the results as they represent the relative variation of a category's prevalence compared to its average: they take on values higher (lower) than zero when the original value is higher (lower) than the average.

\subsection*{Other behavioral markers} \label{sec:validation}

To assess the validity of our operationalization of Strong's model, we compared its results with the output of alternative state-of-the-art text-mining techniques, and with real-world mobility patterns.

\subsubsection*{Interaction types} \label{sec:validation-conversations}

We compared the results obtained via word-matching with a state-of-the-art deep learning tool for Natural Language Processing designed to capture fundamental types of social interactions from conversational language~\citep{choi20social}. This tool uses Long Short-Term Memory neural networks (LSTMs)~\citep{hochreiter1997long} that take in input a 300-dimensional GloVe representation of words~\citep{pennington2014glove} and output a series of confidence scores in the range $[0,1]$ that estimate the likelihood that the text expresses certain types of social interactions. The classifiers exhibited a very high classification performance, up to an Area Under the ROC Curve (AUC) of 0.98. AUC is a performance metric that measures the ability of the model to assign higher confidence scores to positive examples (i.e., text characterized by the type of interaction of interest) than to negative examples, independent of any fixed decision threshold; the expected value for random classification is 0.5, whereas an AUC of 1 indicates a perfect classification.

Out of the ten interaction types that the tool can classify~\citep{deri18coloring}, three were detected frequently with likelihood $>0.5$ in our Twitter data: \emph{conflict} (expressions of contrast or diverging views~\citep{tajfel1979integrative}), social \emph{support} (giving emotional or practical aid and companionship~\citep{fiske2007universal}), and \emph{power} (expressions that mark a person's power over the behavior and outcomes of another~\citep{blau64exchange}). 

Given a tweet's textual message $m$ and an interaction type $i$, we used the classifier to compute the likelihood score $l_i(m)$ that the message contains that interaction type. We then binarized the confidence scores using a threshold-based indicator function: 
\begin{equation}
    l^{\theta}_i(m) = 
    \begin{cases}
      1, & \text{if}\ l_i(m) \geq \theta_i\\
      0, & \text{otherwise}
    \end{cases}
 \end{equation}
Following the original approach~\citep{choi20social}, we used a different threshold for each interaction type, as the distributions of their likelihood scores tend to vary considerably. We thus picked conservatively $\theta_i$ as the value of the $85^{th}$ percentile of the distribution of the confidence scores $l_i$, thus favoring precision over recall. Last, similar to how we constructed temporal signals for the language categories, we counted the number of users $U_i(t)$ who posted at least one tweet at time $t$ that contains interaction type $i$. We then obtained the fraction of users who mentioned interaction type $i$ by dividing $U_i(t)$ by the total number of users $U(t)$ who tweeted at time $t$:
\begin{equation}
f_i(t) = \frac{U_i(t)}{U(t)}.
\label{eqn:10dims}
\end{equation}
Last, we min-max normalized these fractions, considering the minimum and maximum values during the whole time period $[0,T]$:
\begin{equation}
f_i(t) = \frac{f_i(t) - \min_{[0,T]}(f_i)} {\max_{[0,T]}(f_i) - \min_{[0,T]}(f_i)}.
\label{eqn:10dims_normalized}
\end{equation}

\subsubsection*{Mentions of medical entities} \label{sec:validation-medical}

We used a state-of-the-art deep learning method for medical entity extraction to identify medical symptoms on Twitter in relation to COVID-19~\citep{scepanovic2020extracting}. When applied to tweets, the method extracts $n$-grams representing medical symptoms (e.g., ``feeling sick''). This method is based on the Bi-LSTM sequence-tagging architecture~\citep{huang2015bidirectional} in combination with GloVe word embeddings~\citep{pennington2014glove} and RoBERTa contextual embeddings~\citep{liu2019roberta}. To optimize the entity extraction performance on noisy textual data from social media, we trained its sequence-tagging architecture on the Micromed database~\citep{jimeno2015identifying}, a collection of tweets manually labeled with medical entities. The hyper-parameters we used are: $256$ hidden units, a batch size of $4$, and a learning rate of $0.1$ which we gradually halved whenever there was no performance improvement after $3$ epochs. We trained for a maximum of $200$ epochs or before the learning rate became too small ($\leq .0001$). The final model achieved an F1-score of $.72$ on Micromed. The F1-score is a performance measure that combines precision (the fraction of extracted entities that are actually medical entities) and recall (the fraction of medical entities present in the text that the method is able to retrieve). We based our implementation on Flair~\citep{akbik-etal-2019-flair} and Pytorch~\citep{paszke2017automatic}, two popular deep learning libraries in Python.

For each unique medical entity $e$, we counted the number of users $U_e(t)$ who posted at least one tweet at time $t$ that mentioned that entity. We then obtained the fraction of users who mentioned medical entity $e$ by dividing $U_e(t)$ by the total number of users $U(t)$ who tweeted at time $t$:
\begin{equation}
f_e(t) = \frac{U_e(t)}{U(t)}.
\label{eqn:medical_entities}
\end{equation}
Last, we min-max normalize these fractions, considering the minimum and maximum values during the whole time period $[0,T]$:
\begin{equation}
f_e(t)= \frac{f_e(t) - \min_{[0,T]}(f_e)} {\max_{[0,T]}(f_e) - \min_{[0,T]}(f_e)}.
\label{eqn:medical_entities_normalized}
\end{equation}

\subsubsection*{Mobility traces} \label{sec:validation-mobility}

Foursquare is a local search and discovery mobile application that relies on the users' past mobility records to recommend places user might may like. The application uses GPS geo-localization to estimate the user position and to infer the places they visited. In response to the COVID-19 crisis, Foursquare made publicly available the data gathered from a pool of 13 million U.S. users. These users were ``always-on'' during the period of data collection, meaning that they allowed the application to gather geo-location data at all times, even when the application was not in use. The data (published through the \url{visitdata.org} website) consists of the daily number of users $v_{s,j}$ visiting any venue of type $j$ in state $s$, starting from February $1^{st}$ to the present day (e.g., 419,256 users visited schools in Indiana on February $1^{st}$). Overall, 35 distinct location categories were provided. To obtain country-wide temporal indicators, we first applied a min-max normalization to the $v_{s,j}$ values:
\begin{equation}
v_{s,j}(t) = \frac{v_{s,j}(t) - \min_{[0,T]}(v_{s,j})} {\max_{[0,T]}(v_{s,j}) - \min_{[0,T]}(v_{s,j})}.
\label{eqn:foursquare}
\end{equation}
We then averaged the values across all states:
\begin{equation}
v_{j}(t) =  \frac{1}{S} \sum_{s} v_{s,j}(t),
\label{eqn:foursquare_agg}
\end{equation}
where $S$ is the total number of states. By weighting each state equally, we obtained a measure that is more representative of the whole U.S. territory, rather than being biased towards high-density regions.

\subsection*{Time series smoothing}

All our temporal indicators are affected by large day-to-day fluctuations. To extract more consistent trends out of our time series, we applied a smoothing function---a common practice when analyzing temporal data extracted from social media~\citep{o2010tweets}. Given a time-varying signal $x(t)$, we apply a ``boxcar'' moving average over a window of the previous $k$ days:
\begin{equation}
x^{*}(t) =  \frac{\sum_{i=t-k}^t{x(i)}}{k};
\label{eqn:smoothing}
\end{equation}
We selected a window of one week ($k=7$). Weekly time windows are typically used to smooth out both day-to-day variations as well as weekly periodicities~\citep{o2010tweets}. We applied the smoothing to all the time series: the language categories ($z^*_c(t)$), the mentions of medical entities ($f^*_e(t)$), the interaction types ($f^*_i(t)$), and the foursquare visits ($v^*_{j}(t)$).

\subsection*{Change-point detection}

To identify phases characterized by different combinations of the language categories, we identified \emph{change-points}---periods in which the values of all categories varied considerably at once. To quantify such variations, for each language category $c$, we computed $\nabla (z^{*}_c(t))$, namely the daily average \emph{squared gradient}~\citep{lutkepohl2005new} of the smoothed standardized fractions of that category. To calculate the gradient, we used the Python function \texttt{numpy.gradient}. The gradient provides a measure of the rate of increase or decrease of the signal; we consider the absolute value of the gradient, to account for the magnitude of change rather than the direction of change. To identify periods of consistent change as opposed to quick instantaneous shifts, we apply temporal smoothing (Equation.~\ref{eqn:smoothing}) also to the time-series of gradients, and we denote the smoothed squared gradients with $\nabla^{*}$. Last, we average the gradients of all language categories to obtain the overall gradient over time:
\begin{equation}
\nabla(t) = \frac{1}{D} \sum_d \nabla^{*} (z^{*}_c(t)).
\label{eqn:gradient}
\end{equation}
Peaks in the time series $\nabla(t)$ represent the days of highest variation, and we marked them as change-points. Using the Python function \texttt{scipy.signal.find\_peaks}, we identified peaks as the local maxima whose values is higher than the average plus two standard deviations, as it is common practice~\citep{palshikar2009simple}.

%%%%%%%%%%%%%%%%%%%%%%%%%%%%%%%%%%%%%%%
%%%%%%%%%%%%%%%%%%%%%%%%%%%%%%%%%%%%%%%
\section*{Results}
%%%%%%%%%%%%%%%%%%%%%%%%%%%%%%%%%%%%%%%
%%%%%%%%%%%%%%%%%%%%%%%%%%%%%%%%%%%%%%%

\subsection*{Language use until the first contagion wave.}

\begin{figure*}[h!]
    \centering
		\includegraphics[width=0.95\linewidth]{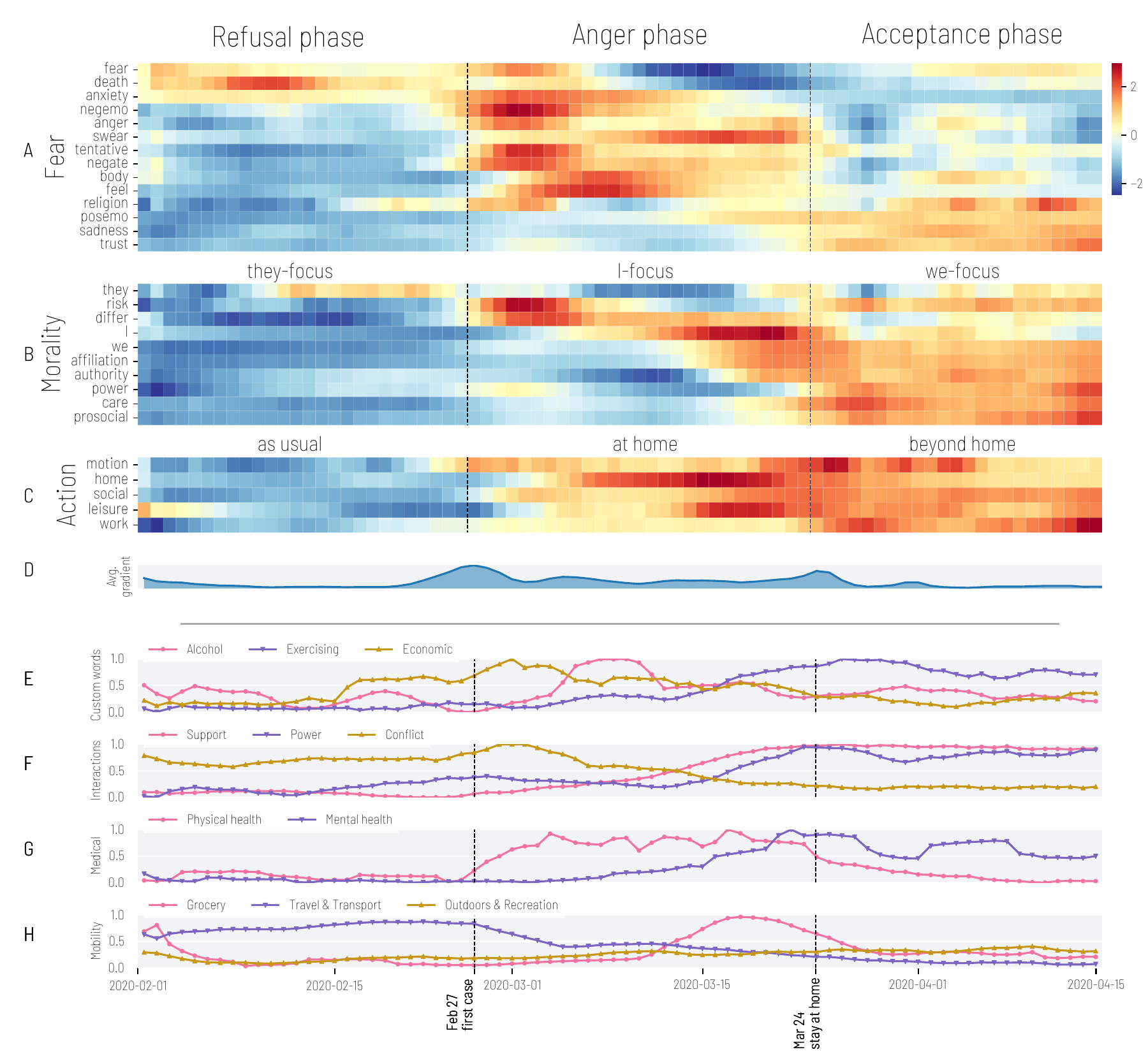}
		\captionsetup{width=.99\linewidth}
		\caption{The Epidemic Psychology on Twitter during the first contagion wave. \emph{(A-C)} evolution of the use of different language categories over time in tweets related to COVID-19. Each row in the heatmaps represents a language category (e.g., words expressing anxiety) that our manual coding associated with one of the three psycho-social epidemics. The cell color represents the daily standardized fraction of people who used words related to that category: values that are higher than the average are red and those that are lower are blue. Categories are partitioned in three groups according to the type of psycho-social epidemics they model: Fear, Morality, and Action. \emph{(D)} average gradient (i.e., instantaneous variation) of all the language categories; the peaks of gradient identify \emph{change-points}---dates around which a considerable change in the use of multiple language categories happened at once. The dashed vertical lines that cross all the plots represent these change-points. \emph{(E-H)} temporal evolution of four families of indicators we used to corroborate the validity of the trends identified by the language categories. We checked internal validity by comparing the language categories with a custom keyword-search approach and two deep-learning NLP tools that extract types of social interactions and mentions of medical symptoms. We checked external validity by looking at mobility patterns in different venue categories as estimated by the GPS geo-localization service of the Foursquare mobile app. The timeline at the bottom of the figure marks some of the key events of the COVID-19 pandemic in the U.S. such as the announcements of the first infection of COVID-19 recorded.}
    \label{fig:heatmap1}
    \vspace{10pt}
\end{figure*}

During the first wave, U.S. residents experienced what the pandemic entailed for the first time, and did so by going through an entire life-cycle, making this contagion wave self-contained:
arrival of an unknown virus, skepticism, isolation measures, full lock-down, and first reopening. Figure~\ref{fig:heatmap1}~(A-C) shows how the standardized fractions of all the language categories as per Formula~(\ref{eqn:z_d_t}) changed from February $1^{st}$ to April $15^{th}$, the day in which restrictions in most states were lifted. The cell color encodes values higher than the average in red, and lower in blue. We partitioned the language categories according to the three psycho-social epidemics. Figure~\ref{fig:heatmap1}~D shows the value of the average squared gradient over time (as per Formula \ref{eqn:gradient}); peaks in the curve represent the days of high local variation. We marked the peaks above two standard deviations from the mean as change-points. We found two change-points that coincide with two key events: February $27^{th}$, the day of the announcement of the first infection in the country; and March $24^{th}$, the day of the announcement of the `stay at home' orders. These change-points identify three phases, which are described next by dwelling on the \emph{peaks} of the different language categories (days when their standardized fractions reached the maximum) and reporting the percentage increase at peak (the increase is compared to the average from February $1^{st}$ to April $15^{th}$, and its peak is denoted by `max peak' in Table~\ref{table:literature_review}). The first phase (\emph{refusal phase}) was characterized by \emph{anxiety} and \emph{fear}. \emph{Death} was frequently mentioned, with a peak on February 11 of +45\% compared to its average during the whole time period. The pronoun \emph{they} was used in this temporal state more than average; this suggests that the focus of discussion was on the implications of the viral epidemic on `others', as this was when no infection had been discovered in the U.S. yet. All other language categories exhibited no significant variations, which reflected an overall situation of `business-as-usual.'

The second phase (\emph{anger phase}) began on February $27^{th}$ with an outburst of \emph{negative emotions} (predominantly \emph{anger}), right after the first COVID-19 contagion in the U.S. was announced. The abstract fear of death was replaced by expressions of concrete health concerns, such as words expressing \emph{risk}, and mentions of how \emph{body} parts did \emph{feel}. On March $13^{th}$, the federal government announced the state of national emergency, followed by the enforcement of state-level `stay at home' orders. During those days, we observed a sharp increase of the use of the pronoun \emph{I} and of \emph{swear} words (with a peak of +54\% on March $18^{th}$), which hints at a climate of discussion characterized by conflict and polarization. At the same time, we observed an increase in the use of words related to the daily habits affected by the impending restriction policies, such as \emph{motion}, \emph{social} activities, and \emph{leisure}. The mentions of words related to \emph{home} peaked on March $16^{th}$ (+38\%), the day when the federal government announced social distancing guidelines to be in place for at least two weeks.

The third phase (\emph{acceptance phase}) started on March $24^{th}$, the day after the first physical-distancing measures were imposed by law. The increased use of words of \emph{power} and \emph{authority} likely reflected the emergence of discussion around the new policies enforced by government officials and public agencies. As the death toll raised steadily---hitting the mark of 1,000 deaths on March $26^{th}$---expressions of conflict faded away, and words of \emph{sadness} became predominant. In those days of hardship, a sentiment of \emph{care} for others and expressions of \emph{prosocial} behavior became more frequent (+19\% and +25\%, respectively). Last, mentions of \emph{work}-related activities peaked as many people either lost their job, or were compelled to work from home as result of the lockdown.

\subsection*{Thematic analysis}

The language categories capture broad concepts related to Strong's epidemic psychology theory, but they do not allow for an analysis of the fine-grained topics within each category. To study them, for each of the 87 combinations of language category and phase (29 language categories, for 3 phases), we listed the 100 most retweeted tweets (e.g., most popular tweets containing anxiety posted in the refusal phase). To identify overarching \emph{themes}, we followed two steps that are commonly adopted in thematic analysis~\citep{braun2006using,smith2012interpretative}. We first applied \emph{open coding} to identify key concepts that emerged across multiple tweets; specifically, one of the authors read all the tweets and marked them with keywords that reflected the key concepts expressed in the text. We then used \emph{axial coding} to identify relationships between the most frequent keywords to summarize them in semantically cohesive themes. Themes were reviewed in a recursive manner rather than linear, by re-evaluating and adjusting them as new tweets were parsed. Table~\ref{table:thematic} summarizes the most recurring themes, together with some of their representative tweets. In the refusal phase, statements of skepticism were re-tweeted widely (Table~\ref{table:thematic}, row 1). The epidemic was frequently depicted as a ``foreign'' problem (r. 2) and all activities kept business as usual (r. 3). 

In the anger phase, the discussion was characterized by outrage against three main categories: foreigners---especially Chinese individuals, as Supplementary Materials detail---(r. 4), political opponents (r. 5), and people who adopted different behavioral responses to the outbreak (r. 6). This level of conflict corroborates Strong's postulate of the \emph{``war against each other''}. Science and religion were two prominent topics of discussion. A lively debate raged around the validity of scientists' recommendations (r. 7). Some social groups put their hopes on God rather than on science (r. 8). Mentions of people self-isolating at home became very frequent, and highlighted the contrast between judicious individuals and careless crowds (r. 9).

Finally, during the acceptance phase, the outburst of anger gave in to the sorrow caused by the mourning of thousands of people (r. 10). By accepting the real threat of the virus, our Twitter users were more open to find collective solutions to the problem and overcome fear with hope (r. 11). Although the positive attitude towards the authorities seemed prevalent, some users expressed disappointment against the restrictions imposed (r. 12). Those who were isolated at home started imagining a life beyond the isolation, especially in relation to reopening businesses (r. 13).

{\def\arraystretch{1.5}
\begin{table}[h!]
\centering
\footnotesize
\begin{tabular}{p{3mm} p{25mm} p{120mm}}
\specialrule{.1em}{.05em}{.05em} 
& \textbf{Theme} & \multicolumn{1}{c}{\textbf{Example tweets}} \\
\Xhline{2\arrayrulewidth}
\multicolumn{3}{c}{\emph{\textbf{The refusal phase}}}  \\
1 &  denial & \emph{``Less than 2\% of all cases result in death. Approximately equivalent to seasonal flu. Relax people.''} \\ 
2 & they-focus & \emph{``We will continue to call it the \#WuhanVirus, which is exactly what it is.''} \\ 
3 & business as usual & \emph{``Agriculture specialists at Dulles airport continue to protect our nation's vital agricultural resources.''} \\

\hline
\multicolumn{3}{c}{\emph{\textbf{The anger phase}}} \\

4 & anger \emph{vs.} foreigners & \emph{``Is there anything you won't use to stir up hatred against the foreigner? \#COVID19 is a global pandemic.''} \\
5 & anger \emph{vs.} political opponents & \emph{``A new level of sickness has entered the body politic. The son of the monster mouthing off grotesque lies about Dems cheering \#coronavirus and Wall Street crashing because we want an end to his father's winning streak.''} \\
6 &  anger \emph{vs.} each other & \emph{``Coronavirus or not, if you are ill, stay the f**k home. You're not a hero for going to work when you are unwell.''} \\
7 & science debate & \emph{``When it comes to how to fight \#CoronavirusPandemic, I'm making my decisions based on healthcare professionals like Dr. Fauci and others, not political punditry''} \\
8 & religion & \emph{``no problem is too big for God to handle [...] with God's help, we will overcome this threat.''} \\
9 & I-focus, home & \emph{``People get upset and annoyed at me when I tweet about the coronavirus, when I urge people to stay in and avoid crowds''}, \emph{``I am in the high risk category for coronavirus so do me a favor [...] beg others to stay at home''} \\

\hline
\multicolumn{3}{c}{\emph{\textbf{The acceptance phase}}} \\

10 & sadness & \emph{``We deeply mourn the 758 New Yorkers we lost yesterday to COVID-19. New York is not numb. We know this is not just a number---it is real lives lost forever.''} \\
11 & we-focus, hope & \emph{``We are thankful for Japan's friendship and cooperation as we stand together to defeat the \#COVID19 pandemic.'', ``During tough times, real friends stick together. The U.S. is thankful to \#Taiwan for donating 2 million face masks to support our healthcare '', ``Now more than ever, we need to choose hope over fear. We will beat COVID-19. We will overcome this. Together.''} \\
12 & authority & \emph{``You can't go to church, buy seeds or paint, operate your business, run on a beach, or take your kids to the park. You do have to obey all new `laws', wear face masks in public, pay your taxes. Hopefully this is over by the 4th of July so we can celebrate our freedom.} \\
13 & resuming work & \emph{``We need to help as many working families and small businesses as possible. Workers who have lost their jobs or seen their hours slashed and families who are struggling to pay rent and put food on the table need help immediately. There's no time to waste.''} \\

\specialrule{.1em}{.05em}{.05em} 
\end{tabular}
\captionsetup{width=.99\linewidth}
\caption{Recurring themes in the three phases, found by the means of thematic analysis of tweets. Themes are paired with examples of popular tweets.}
\label{table:thematic}
\end{table}
}

\subsection*{Comparison with other behavioral markers}

To assess the validity of our approach, we compared the previous results with the output of alternative text-mining techniques applied to the same data (\emph{internal validity}), and with real-world mobility traces (\emph{external validity}).

\subsubsection*{Comparison with other text mining techniques }

We processed the very same social media posts with three alternative text-mining techniques (Figure~\ref{fig:heatmap1} E-G). In Table~\ref{table:correlations}, we reported the three language categories with the strongest correlations with each behavioral marker. 

First, to allow for interpretable and explainable results, we applied a simple word-matching method that relies on a custom lexicon containing three categories of words reflecting consumption of alcohol, physical exercising, and economic concerns, as those aspects have been found to characterize the COVID-19 pandemic~\citep{economist-article}. We measured the daily fraction of users mentioning words in each of those categories (Figure~\ref{fig:heatmap1} E). In the refusal phase, the frequency of any of these words did not significantly increase. In the anger phase, the frequency of words related to economy peaked, and that related to alcohol consumption peaked shortly after that. Table~\ref{table:correlations} shows that economy-related words were highly correlated with the use of \emph{anxiety} words ($r=0.73$), which is in line with studies indicating that the degree of apprehension for the declining economy was comparable to that of health-hazard concerns~\citep{fetzer2020coronavirus,bareket2020covid}. Words of alcohol consumption were most correlated with the language dimensions of \emph{body} ($r=0.70$), \emph{feel} ($r=0.62$), \emph{home} ($r=0.58$); in the period were health concerns were at their peak, home isolation caused a rising tide of alcohol use~\citep{da2020covid,finlay2020covid}. Finally, in the acceptance phase, the frequency of words related to physical exercise was significant; this happened at the same time when the use of positive words expressing togetherness was at its highest---\emph{affiliation} ($r=0.95$), \emph{posemo} ($r=0.93$), \emph{we} ($r=0.92$). All these results match our previous interpretations of the peaks for our language categories.

Second, since it is unclear whether a simple word count approach is effective in studying how the three psycho-social epidemics unfolded over time, we additionally applied the deep-learning approach that extracts mentions of expressions of \emph{conflict}, \emph{social support}, and \emph{power}. Figure~\ref{fig:heatmap1}~F shows the min-max normalized scores of the fraction of users posting tweets labeled with each of these three interaction types (as per Formula~\ref{eqn:10dims_normalized}). In the refusal phase, \emph{conflict} increased---this is when anxiety and blaming foreigners were recurring themes in Twitter. In the anger phase, conflict peaked (similar to \emph{anxiety} words, $r=0.88$), yet, since the first lock-down measures were announced, initial expressions of power and of social support gradually increased as well. Finally, in the acceptance phase, social support peaked. Support was most correlated with the categories of \emph{affiliation} ($r=0.98$), \emph{positive emotions} ($r=0.96$), and \emph{we} ($r=0.94$) (Table~\ref{table:correlations}); power was most correlated with \emph{prosocial} ($r=0.95$), \emph{care} ($r=0.94$), and \emph{authority} ($r=0.94$). Again, our previous interpretations concerning the existence of a phase of conflict followed by a phase of social support were further confirmed by the deep-learning tool, which, as opposed to our dictionary-based approaches, does not rely on word matching. 

Third, we used the deep-learning tool that extracts mentions of medical entities from text~\citep{scepanovic2020extracting}. Out of all the entities extracted, we focused on the 100 most frequently mentioned and grouped them into two families of symptoms, respectively, those related to physical health (e.g., ``fever'', ``cough'', ``sick'') and those related to mental health (e.g., ``depression'', ``stress'')~\citep{brooks2020psychological}. The min-max normalized fractions of users posting tweets containing mentions of these symptoms (as per Formula~\ref{eqn:medical_entities_normalized}) are shown in Figure~\ref{fig:heatmap1}~G. In refusal phase, the frequency of symptom mentions did not change. In the anger phase, instead, physical symptoms started to be mentioned, and they were correlated with the language categories expressing panic and physical health concerns---\emph{swear} ($r=0.83$), \emph{feel} ($r=0.77$), and \emph{negate} ($r=0.67$). In the acceptance phase, mentions of mental symptoms became most frequent. Interestingly, mental symptoms peaked when the Twitter discourse was characterized by positive feelings and prosocial interactions---\emph{affiliation} ($r=0.91$), \emph{we} ($r=0.88$), and \emph{posemo} ($r=0.85$); this is in line with recent studies that found that the psychological toll of COVID-19 has similar traits to post-traumatic stress disorders and its symptoms might lag several weeks from the period of initial panic and forced isolation~\citep{galea2020mental,liang2020effect,dutheil2020ptsd}.

\subsubsection*{Comparison with mobility traces}

To test for the external validity of our language categories, we compared their temporal trends with the mobility data from Foursquare. We picked three venue categories: \emph{Grocery} shops, \emph{Travel \& Transport}, and \emph{Outdoors \& Recreation} to reflect three different types of fundamental human needs~\citep{maslow1943theory}: the primary need of getting food supplies, the secondary need of moving around freely (or to limit mobility for safety), and the higher-level need of being entertained. In Figure~\ref{fig:heatmap1}~H, we show the min-max normalized number of visits over time (as per Formula~\ref{eqn:foursquare_agg}). The periods of higher variations of the normalized number of visits match the transitions between the three phases. In the refusal phase,  mobility patterns did not change. In the anger phase, instead, travel started to drop, and grocery shopping peaked, supporting the interpretation of a phase characterized by a wave of panic-induced stockpiling and a compulsion to save oneself---it co-occurred with the peak of use of the pronoun \emph{I} ($r=0.80$)---rather than helping others. Finally, in the acceptance phase, the panic around grocery shopping faded away, and the number of visits to parks and outdoor spaces increased.

\subsection*{Embedding epidemic psychology in real-time models}

To embed our operationalization of epidemic psychology into real-time models (e.g., epidemiological models, urban mobility models), our measures need to work at any point in time during a new pandemic; yet, given their current definitions, they do not: that is because they are normalized values over the \emph{whole} period of study (Figure~\ref{fig:heatmap1}~A-C). To fix that, we designed a new composite measure that does not rely on full temporal knowledge, and a corresponding detection method that determines which of the three phases one is in at any given point in time.

{\def\arraystretch{1.5}
\begin{table*}[t!]
\setlength{\tabcolsep}{1.5mm}
\centering
\footnotesize
\begin{tabular}{l | p{21mm}  p{21mm} p{21mm} | p{21mm} p{21mm} p{21mm}}

\specialrule{.1em}{.05em}{.05em} 
\textbf{Phase} & \multicolumn{3}{c}{\textbf{Top positive}} & \multicolumn{3}{c}{\textbf{Top negative}} \\
\specialrule{.1em}{.05em}{.05em} 
Refusal           & \textbf{death} (0.66) & they (0.06) & fear (0.04) & \textbf{I} (-1.51) & we (-1.27) & home (-1.22) \\
Anger & \textbf{swear} (2.17) & feel (1.51) & anxiety (1.46) & \textbf{death} (-0.70) & sadness (-0.51) & prosocial (-0.38) \\
Acceptance        & \textbf{sad} (1.35) & affiliation (1.19) & prosocial (1.17) & \textbf{anxiety} (-1.62)  & swear (-1.36) & I (-0.34) \\
\specialrule{.1em}{.05em}{.05em} 

\end{tabular}
\captionsetup{width=.99\linewidth}
\caption{Top three positive and bottom negative beta coefficients of the logistic regression models for the three phases. The categories in bold are those included in our composite temporal score.}
\label{table:coefficients}
%\vspace{-15pt}
\end{table*}
}

First, for each language category $c$, we computed the average value of $f_c$ (as per Formula~\ref{eqn:f_d_t}) during the first day of the epidemic, specifically $\mu_{[0,1]}(f_c)$. During the first day, 86k users tweeted. We experimented with longer periods (up to a week and 0.4M users), and obtained qualitatively similar results. We used the averages computed on this initial period as reference values for later measurements. The assumption behind this approach is that the modeler would know the set of relevant hashtags in the initial stages of the pandemic, which is reasonable considering that this was the case for all the major pandemics occurred in the last decade~\citep{fu2016people,oyeyemi2014ebola,chew2010pandemics}. Starting from the second day, we then calculated the percent change of the $f_c$ values compared to the reference values:
\begin{equation}
\Delta\%_c(t) = \frac{f_c(t) - \mu_{[0,1]}(f_c)}{\mu_{[0,1]}(f_c)}.
\label{eqn:delta}
\end{equation}
For each phase, we defined a parsimonious measure composed only of two dimensions: the dimension most positively associated with the phase (expressed in percent change) minus that most negatively associated with it (e.g., (\emph{death} - \emph{I}) for the refusal phase). To identify such dimensions, we trained three logistic regression binary classifiers (one per phase). For each phase, we marked with label 1 all the days that were included in that phase and with 0 those that were not. Then, we trained a classifier to estimate the probability $P_{phase_i}(t)$ that day $t$ belongs to phase $i$, out of the $\Delta\%_c(t)$ values for all categories. During training, the logistic regression learned coefficients for each of the categories. On average, the classifiers were able to identify the correct phase for 98\% of the days.

The regressions coefficients were then used to rank the language category by their predictive power. Table~\ref{table:coefficients} shows the top three positive beta coefficients and bottom three negative ones for each of the three phases. The top and bottom categories of all phases belong to the LIWC lexicon. For each phase, we subtracted the top category from the bottom category without considering their beta coefficients, as these would require, again, full temporal knowledge:
\begin{align}
\begin{split}
& \Delta\%_{Refusal}            =  \Delta\%_{death} - \Delta\%_{I}  \\
& \Delta\%_{Anger}  =  \Delta\%_{swear} - \Delta\%_{death} \\
& \Delta\%_{Acceptance}         =  \Delta\%_{sad} - \Delta\%_{anxiety}
\end{split}
\label{eqn:delta3-3}
\end{align}
The resulting composite measure has the same change-points (Figure~\ref{fig:stackplot}A) as the full-knowledge measure's (Figure~\ref{fig:heatmap1}), suggesting that the real-time and parsimonious computation does not compromise the original trends. In a real-time scenario, transition between phases are captured by the changes of the dominant measure; for example, when the \emph{refusal} curve is overtaken by the \emph{anger} curve. In addition, we correlated the composite measures with each of the behavioral markers we used for validation (Figure~\ref{fig:heatmap1} E-H) to find which are the markers most typically associated with each of the phases. We reported the correlations in Table~\ref{table:correlations}. During the refusal phase, conflictual interactions were frequent ($r=0.58$) and long-range mobility was common ($r=0.62$); during the anger phase, as mobility reduced~\citep{engle2020staying,gao2020mapping}, some people hoarded groceries and alcohol~\citep{da2020covid,finlay2020covid} and expressed concerns for their physical health ($r=0.81$) and for the economy~\citep{fetzer2020coronavirus,bareket2020covid}; last, during the acceptance phase, some people ventured outdoors, started exercising more, and expressed a stronger will to support each other ($r=0.90$), in the wake of a rising tide of deaths and mental health symptoms ($r=0.85$)~\citep{galea2020mental,liang2020effect,dutheil2020ptsd}.

{\def\arraystretch{2}
\begin{table*}[t!]
\setlength{\tabcolsep}{1.5mm}
\centering
\footnotesize
\begin{tabular}{l l p{23mm} p{23mm} p{23mm} | c cc}

\specialrule{.1em}{.05em}{.05em} 
& & & & & \multicolumn{3}{c}{\textbf{Correlation with phases}} \\
& \textbf{Marker} & \multicolumn{3}{c|}{\textbf{Most correlated language categories}} & Refusal & \multicolumn{1}{c}{Anger} & Acceptance  \\
\specialrule{.1em}{.05em}{.05em} 
\multirow{3}{*}{\rotatebox[origin=c]{90}{\textbf{Custom words}}} &  Alcohol & body (0.70) & feel (0.62) & home (0.58) & -0.43 & \textbf{0.46} & -0.12  \\
& Economic & anxiety (0.73) & negemo (0.68) & negate (0.56)  & -0.12 & \textbf{0.37} & -0.53 \\
& Exercising & affiliation (0.95) & posemo (0.93) & we (0.92) & -0.62 & 0.31 & \textbf{0.89}  \\
\hline
\multirow{3}{*}{\rotatebox[origin=c]{90}{\textbf{Interactions}}} 
& Conflict & anxiety (0.88) & death (0.57) & negemo (0.54)  & \textbf{0.58} & -0.24 & -0.92  \\
& Support & affiliation (0.98) & posemo (0.96) & we (0.94) & -0.68 & 0.37 & \textbf{0.90}   \\
& Power & prosocial (0.95) & care (0.94) & authority (0.94)  & -0.48 & 0.18 & \textbf{0.88}  \\
\hline
\multirow{2}{*}{\rotatebox[origin=c]{90}{\textbf{Medical}}} &   Physical health & swear (0.83) & feel (0.77) & negate (0.67) & -0.66 & \textbf{0.81} & -0.32 \\
& Mental health & affiliation (0.91) & we (0.88) & posemo (0.85)  & -0.65 & 0.36 & \textbf{0.85}  \\
\hline
\multirow{3}{*}{\rotatebox[origin=c]{90}{\textbf{Mobility}}} 
& Travel & death (0.59) & anxiety (0.58) & & \textbf{0.62} & -0.32  & -0.82 \\
&  Grocery & I (0.80) & leisure (0.72) & home (0.64)  & -0.77 & \textbf{0.70} & 0.29  \\
& Outdoors & sad (0.68) & posemo (0.65) & affiliation (0.59)  & -0.62 & 0.39 & \textbf{0.72}  \\
\specialrule{.1em}{.05em}{.05em} 

\end{tabular}
\captionsetup{width=.99\linewidth}
\caption{(Left) Correlation of our language categories with behavioral markers computed with alternative techniques and datasets. For each marker, the three categories with strongest correlations are reported, together with their Pearson correlation values in parenthesis. (Right) Pearson correlation between values for our behavioral markers and ``being'' in a given phase or not. Values in bold indicate the highest values for each marker  across the three phases. All reported correlations are statistically significant ($p<0.01$).}
\label{table:correlations}
\end{table*}
}

\subsection*{Language use after the first contagion wave.} 
\begin{figure*}[t!]
    \centering
		\includegraphics[width=0.95\linewidth]{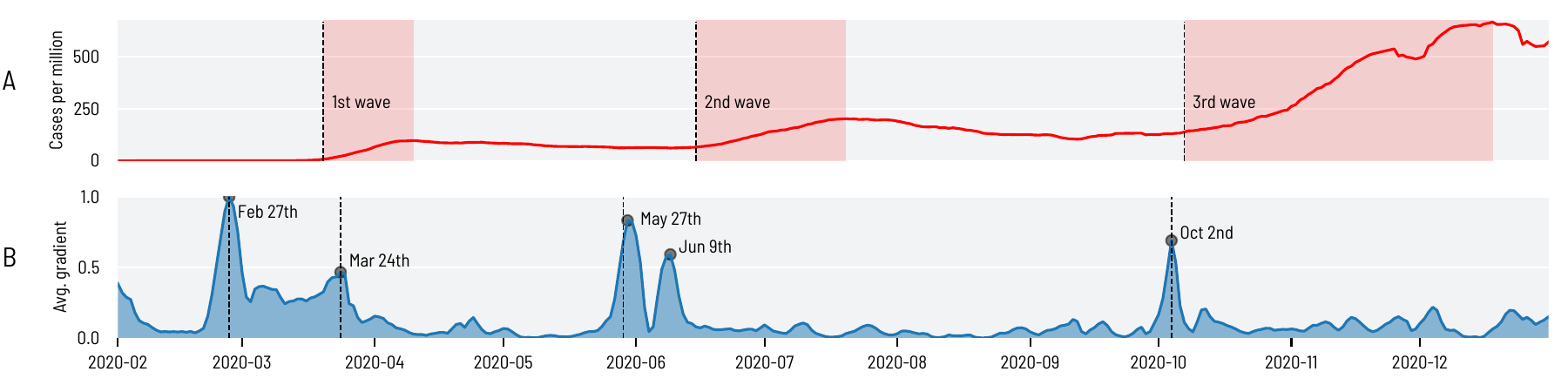}
		\captionsetup{width=.99\linewidth}
		\caption{The number of recorded infections in the U.S. (A), and the average gradient (i.e., instantaneous variation) of all the language categories (B). There were three contagion waves in the year (shaded areas in A) and, for each wave, there were peaks in the gradient (marked with circles in B), which identify \emph{change-points}, that is, periods in which the use of language  considerably changed.}
    \label{fig:timeline}
\end{figure*}

\begin{figure*}[t!]
    \centering
		\includegraphics[width=0.90\linewidth]{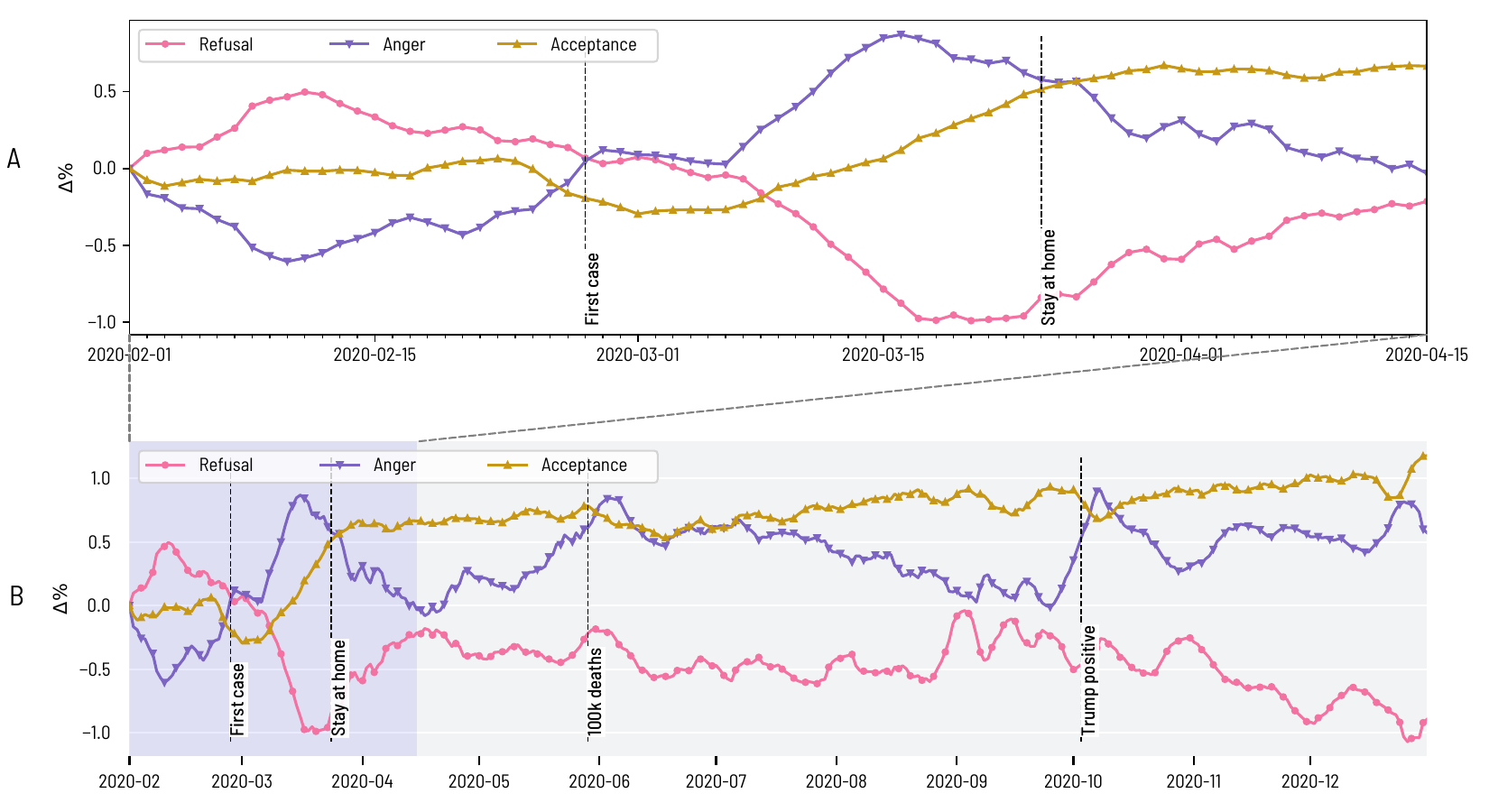}
		\captionsetup{width=.99\linewidth}
		\caption{The evolution of the language categories associated with \emph{refusal}, of those associated with  \emph{anger}, and of those associated with  \emph{acceptance}, zooming out from  the first contagion wave (A) to the three waves during the entire 2020 (B).}
    \label{fig:stackplot}
\end{figure*}

\begin{figure*}[t!]
    \centering
		\includegraphics[width=0.95\linewidth]{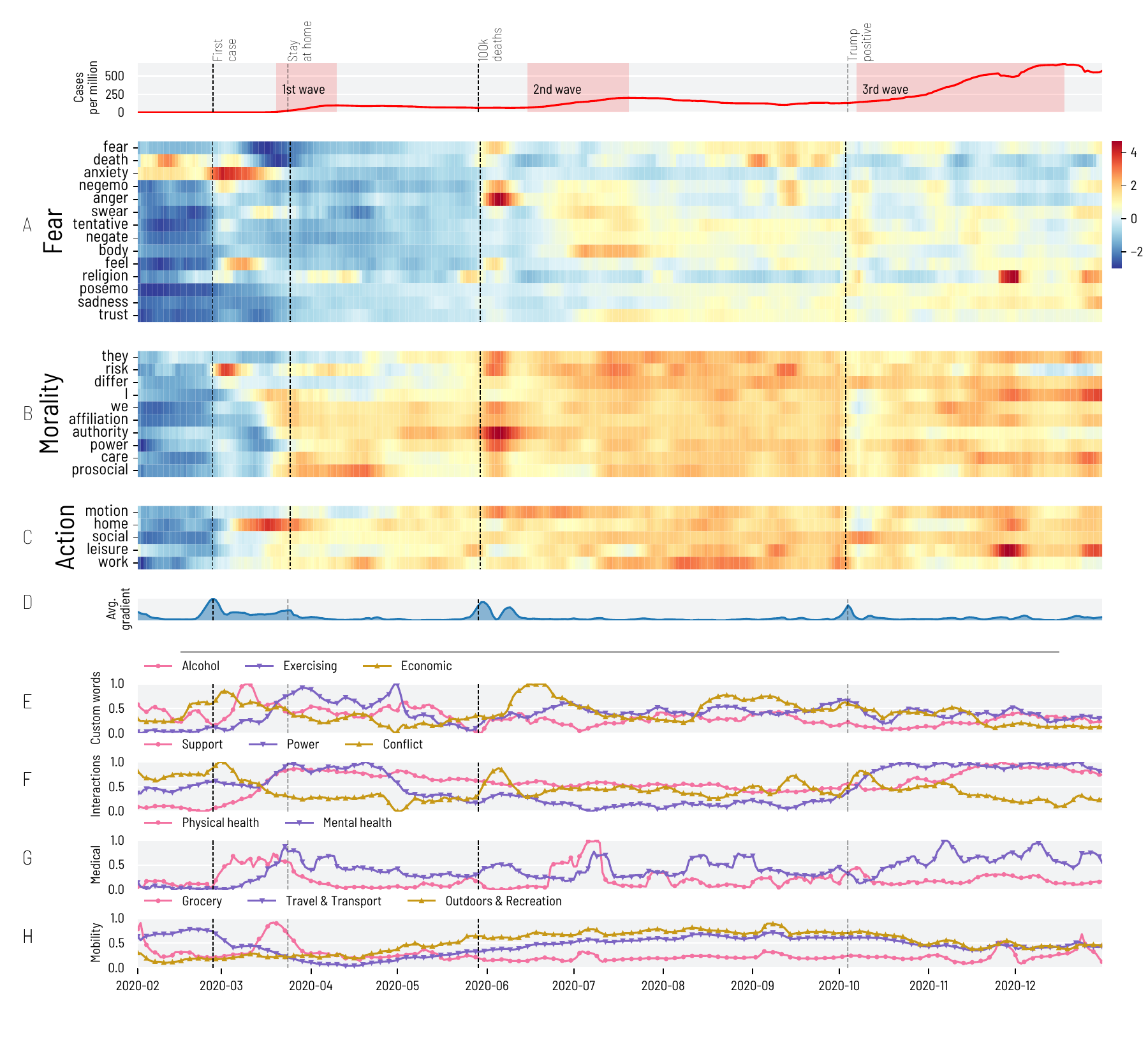}
		\captionsetup{width=.99\linewidth}
		\caption{The Epidemic Psychology on Twitter during the three contagion waves (shown in the top most panel) during the entire year of 2020. (A-H) temporally expands  (A-H)  in Figure~\ref{fig:heatmap1} with the only difference that the heatmaps in (A-C) here show values that were standardized using the mean and standard deviation calculated on the whole period of study, from February $1^{st}$ to December $31^{st}$.}
    \label{fig:heatmap_fulltimeline}
\end{figure*}

After the first wave and the ``stay-at-home'' lock-down at the end of March, for the remaining part of the year, there were two other contagion waves (Figure~\ref{fig:timeline}A): one at the beginning of June, and the other at the beginning of October. In a way similar to   the first wave, these two others were both associated with significant changes in the use of language (Figure~\ref{fig:timeline}B): we had two change peaks for the second wave (i.e., May $27^{th}$ and June $9^{th}$), and one for the third (October 2). This comes at no surprise as these two periods corresponded to two widely discussed events: the first 100K deaths in the U.S., and President Donald Trump testing positive for COVID-19. In particular, Figure~\ref{fig:stackplot}B shows that these changes were both to do with the rumping up of the categories associated with the \emph{anger phase}: discussions of changes in \emph{mobility}, and posts characterized by \emph{anger} and, more generally, by \emph{negative emotions} were predominant. By contrast, the categories associated with  \emph{refusal} and \emph{acceptance} diverged from each other: unsurprisingly, throughout the year, \emph{refusal} gradually died off, while \emph{acceptance} increasingly took hold. Overall, we observed two classes of pattern (Figure~\ref{fig:stackplot}B). First, the three phases were not always orthogonal but blended together at times. During the second contagion wave, for example, \emph{anger} and \emph{acceptance} were both predominant, and had been so for several months. Second, the use of language had a cyclical nature. During each contagion wave, three consecutive local peaks (local maxima) were observed: \emph{refusal}  first, then \emph{anger},  finally \emph{acceptance}. This can be observed for all the three contagion waves. Such a cyclical nature was also reflected in our behavioral markers (Figure~\ref{fig:heatmap_fulltimeline}(E-H)). For each  wave, mentions of \emph{conflict} peaked to then be followed by mentions of \emph{support} and \emph{power} (Figure~\ref{fig:heatmap_fulltimeline}F). The same went for medical conditions: mentions of \emph{physical health} peaked to then be followed by mentions of \emph{mental health} (Figure~\ref{fig:heatmap_fulltimeline}G).

%%%%%%%%%%%%%%%%%%%%%%%%%%%%%%%%%%%%%%%
%%%%%%%%%%%%%%%%%%%%%%%%%%%%%%%%%%%%%%%
\section*{Discussion}
%%%%%%%%%%%%%%%%%%%%%%%%%%%%%%%%%%%%%%%
%%%%%%%%%%%%%%%%%%%%%%%%%%%%%%%%%%%%%%%

\subsection*{Findings beyond Strong's model}

Strong's theory offers a framework upon which to operationalize the three psycho-social epidemics from the use of language but does not specifically describe how these epidemics unfold. Based on our data-driven results, we confirmed that the three epidemics are indeed present in our social conversations spanning almost one year, and they unfolded in  ways that allowed us to enrich what Strong  initially hypothesized in relation to  four main aspects.

First, Strong's theory predicts the presence of the three psycho-social epidemics but does not describe how they would be related to each other over time. We found that the three epidemics, as expressed by the relative presence of their relevant language categories, simultaneously raised and fell over time, and did so in  relation to one another, demarcating three specific temporal phases. Over time, we identified three specific combinations of the epidemics that  generated three phases our  Twitter users went through: an initial \emph{refusal} phase,  an \emph{anger} phase, and a final \emph{acceptance} phase. Since these  temporal phases partly resemble the stages of grief by~\cite{kubler1972death}, a promising direction for future work is to explore the relationship between these phases and the stages of grief.

Second, Strong's narration slightly hints at a typical sequence of events according to which the epidemic of \emph{fear} activates before that of \emph{morality}, which is then followed by that of \emph{action}. We indeed  observed a similar set of events yet these events were not  strictly sequential but rather cyclical. Interestingly, every new cycle started  in conjunction with the same specific event: the diffusion rate of the virus reaching a local maximum. Shortly after every sharp increase in the diffusion rate,  a cycle of refusal, anger, and acceptance unfolded among our Twitter users.

Third, Strong's framework does not make any explicit distinction between the initial stages of an epidemic and its final stages. We found two regimes  that considerably separated the initial stages from the later stages: in the initial cycle, the variation of the three epidemics was of a  larger magnitude than  those  in the two subsequent cycles. 

Last, Strong's theory should be used as a descriptive framework and, as such, cannot be used to  explore  short-term variations. By contrast, a data-driven experimental work such as ours is able to point out when such variations potentially took place, as Figure~\ref{fig:heatmap_fulltimeline} shows. In future work,  these points of change could be subject to qualitative inquiry, which might well enrich the original formulation of the theory.

\subsection*{Implications}

New infectious diseases break out abruptly, and public health agencies try to rely on detailed planning yet often find themselves to improvise around their playbook. They are constantly confronting not only the health epidemic but also the three psycho-social epidemics. Measuring the effects of epidemics on societal dynamics and population mental health has been an open research problem for a long time, and multidisciplinary approaches have been called for~\citep{holmes2020multidisciplinary}. We contributed to this line of research by operationalizing Strong's model and successfully testing it on Twitter in the U.S.. Since our methodology can be applied to any textual data, future work may well study alternative (even cross-cultural) population segments. Since our language categories are not tailored to a specific epidemic (e.g., they do not reflect any specific symptom an epidemic is associated with), our approach can be applied to a future epidemic, provided that the set of relevant hashtags associated with the epidemic is known; this is a reasonable assumption to make though, considering that the consensus on Twitter hashtags is reached quickly~\citep{baronchelli2018emergence}, and that several epidemics that occurred in the last decade sparked discussions on Twitter since their early days~\citep{fu2016people,oyeyemi2014ebola,chew2010pandemics}.

Our method complements the numerous cross-sectional studies on the psychological impact of health epidemics conducted on representative population samples~\citep{shultz20152014,brooks2020psychological}, 
not least because it collects real-time statistics of implicit behavioral signals, which are orthogonal to survey responses.

For computer science researchers, our method could provide a starting point for developing more sophisticated tools for monitoring psycho-social epidemics. Furthermore, from the theoretical standpoint, our work provides the first operationalization of Strong's model of the epidemic psychology, and widens its theoretical implications by observing cyclical phases of diffusion of the psycho-social epidemics.

Finally, the ability to systematically characterize the three psycho-social epidemics from the use of language on social media makes it possible to embed epidemic psychology into models currently used to tackle epidemics such as mobility models~\citep{bansal2016big}. To see how, consider that, in digital epidemiology~\citep{salathe2012digital,bauch2013social}, some parameters of epidemic models are initialized or adjusted based on a variety of digital data to account for co-determinants of the spreading process that are hard to quantify with traditional data sources, especially in the first stages of the outbreak. This is particularly useful when modeling social and psychological processes such as risk perception~\citep{bagnoli2007risk,moinet2018effect}. Interestingly, these approaches are designed to deal with partial data, therefore they can benefit even from digital data that is incomplete and---like in the case of our Twitter-based study---not necessarily representative of the whole population~\citep{salathe2012digital}.

\subsection*{Limitations}

Future work could improve our work in five main aspects. First, we focused only on one viral epidemic, without being able to compare it to others. Yet, if one were to obtain past social media data during the outbreaks of diseases like Zika~\citep{fu2016people}, Ebola~\citep{oyeyemi2014ebola}, and the H1N1 influenza~\citep{chew2010pandemics}, one could apply our methodology in those contexts as well, and identify similarities and differences. For example, one could study how mortality rates or speed of spreading influence the representation of Strong's epidemic psychology on social media.  

Second, our geographical focus was the entire United States and, as such, was coarse and limited in scope. In Supplementary Materials, we broke-down the analysis of temporal phases for individual states in the U.S., and observed no substantial differences across states. In the future, one could conduct a more systematic analysis on a finer geographical granularity, relate differences between states to known events (e.g., a governor's decisions, prevalence of cases, media landscape, and residents' cultural traits). In particular, recent studies suggested that the public reaction to COVID-19 varied across the U.S. states depending on their political leaning~\citep{painter2020political,grossman2020political}. One could also apply our methodology to other English-speaking countries, to investigate how cultural dimensions~\citep{hofstede2005cultures} and cross-cultural personality trait variations~\citep{bleidorn2013personality} might influence the three psycho-social epidemics.

Third, the three psycho-social epidemics were not always orthogonal to each other but did blend together at times. Future work could focus on those particular periods of time to determine whether the use either of finer-grained categorizations of language  or of event detection techniques other than our change-point detection~\citep{aiello2013sensing} could disentangle those  periods in theoretically meaningful ways.

Fourth, our study is limited to Twitter, mainly because Twitter is the largest open stream of real-time social media data. The practice of using Twitter as a way of modeling the psychological state of a country carries its own limitations. Despite having a rather high penetration in the U.S. (around 20\% of adults, according to the latest estimates~\citep{twitter_adoption}), its user base is not representative of the general population~\citep{li2013spatial}. Additionally, Twitter is notoriously populated by bots~\citep{ferrara2016rise,varol2017online}, automated accounts that are often used to amplify specific topics or view points. Bots played an important role to steer the discussion on several events of broad public interest~\citep{bessi2016social,broniatowski2018weaponized}, and it is reasonable to expect that they have a role in COVID-related discussions too, as some studies seem to suggest~\citep{yang2020prevalence}. To partly discount their impact, since they tend to have anomalous levels of activity (especially retweeting~\citep{bessi2016social}), we performed two tests. First, we computed all our measures at user-level rather than tweet-level, which counter anomalous levels of activity. Second, we replicated our temporal analysis excluding retweets, and obtained very similar results. In the future, one could attempt to adapt our framework to different sources of online data, for example to web search queries---which have proven useful to identify different phases of the public reactions to the COVID-19 pandemic~\citep{husnayain2020applications}.

Last, as Strong himself acknowledged in his seminal paper: \emph{``any sharp separation between different types of epidemic psychology is a dubious business.''} Our work has operationalized each psycho-social epidemic independently. In the future, modeling the relationships among the three epidemics might identify hitherto hidden emergent properties.

\bibliographystyle{abbrvnat}

\section*{Acknowledgments}
We thank Sarah Konrath, Rosta Farzan, and Licia Capra for their useful feedback on the manuscript. This research was partly supported by the 
EU Grant ``GO GREEN ROUTES'' no. 869764.

%%%%%%%%%%%%%%%%%%%%%%%%%%%%%%%%%%%%%%%
%%%%%%%%%%%%%%%%%%%%%%%%%%%%%%%%%%%%%%%
\section*{Data Availability}
%%%%%%%%%%%%%%%%%%%%%%%%%%%%%%%%%%%%%%%
%%%%%%%%%%%%%%%%%%%%%%%%%%%%%%%%%%%%%%%

The daily aggregates of the measurements are available at: \url{https://doi.org/10.6084/m9.figshare.14892642.v1}. The tweet IDs we used are available at \url{https://github.com/echen102/COVID-19-TweetIDs}. Other datasets and visualizations are available on the project's site \url{http://social-dynamics.net/EpidemicPsychology}.

%%%%%%%%%%%%%%%%%%%%%%%%%%%%%%%%%%%%%%%
%%%%%%%%%%%%%%%%%%%%%%%%%%%%%%%%%%%%%%%
\section*{Competing Interests}
%%%%%%%%%%%%%%%%%%%%%%%%%%%%%%%%%%%%%%%
%%%%%%%%%%%%%%%%%%%%%%%%%%%%%%%%%%%%%%%

The author(s) declare no competing interests.

\newpage

\section*{Supplementary Materials}

%%%%%%%%%%%%%%%%%%%%%%%%%%%%%%%%%%%%%%%
%%%%%%%%%%%%%%%%%%%%%%%%%%%%%%%%%%%%%%%
\subsection*{Spatial representativeness of Twitter data}
%%%%%%%%%%%%%%%%%%%%%%%%%%%%%%%%%%%%%%%
%%%%%%%%%%%%%%%%%%%%%%%%%%%%%%%%%%%%%%%

To make sure that the geographical penetration of our Twitter data reflects the spatial distribution of the U.S. population, we compared the user activity in each state with the census population. Figure~\ref{fig:popplot} shows that both the number of Twitter users and the volume of their tweets correlate very strongly with the census population population estimates in year 2019 at the level of US states (\url{https://www.census.gov}).

\renewcommand{\thefigure}{S1}
\begin{figure}[h!]
	\centering
		\includegraphics[width=0.90\linewidth]{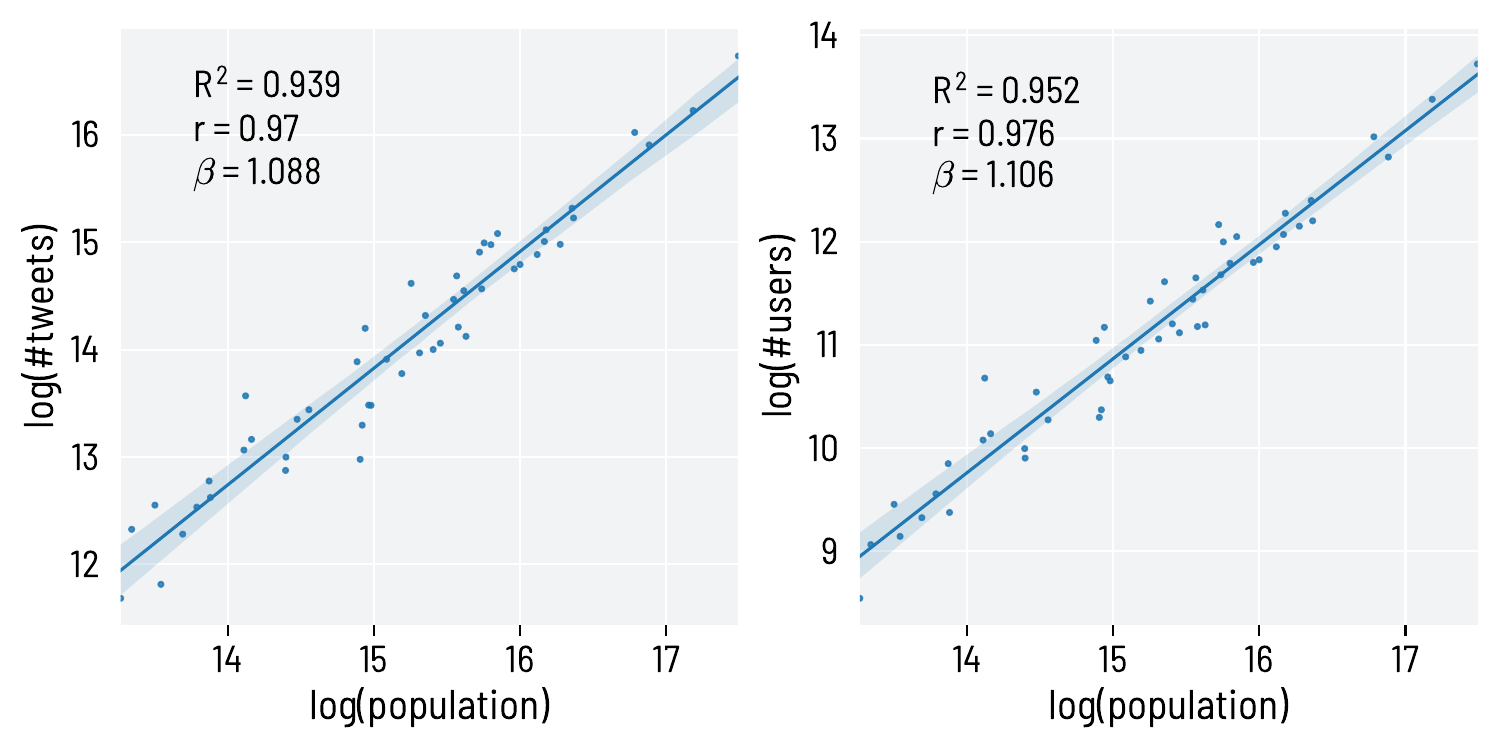}
	\caption{Correlation of the census population of US states estimated in year 2019 with: the number of tweets posted in those states (left), and the number of unique Twitter users who posted at least one tweet (right) during the period of analysis. A linear fit is shown together with the coefficient of determination for a linear regression ($R^2$), the Pearson correlation coefficient ($r$), and the coefficient of the fit ($\beta$). The number of tweets and that of Twitter users scale almost linearly with population estimates.}
	\label{fig:popplot}
\end{figure}

%%%%%%%%%%%%%%%%%%%%%%%%%%%%%%%%%%%%%%%
%%%%%%%%%%%%%%%%%%%%%%%%%%%%%%%%%%%%%%%
\subsection*{Signals of racism in Twitter discussions}
%%%%%%%%%%%%%%%%%%%%%%%%%%%%%%%%%%%%%%%
%%%%%%%%%%%%%%%%%%%%%%%%%%%%%%%%%%%%%%%

To corroborate the qualitative intuition that the first phase was characterized by discussions that depicted the pandemic as a foreign problem, often using racist undertones, we measured the volume of tweets containing hashtags that were clear indicators of content that is either racist or aimed at antagonizing China. We manually selected those hashtags from the list of hashtags that appeared at least 100 times in the dataset. Figure~\ref{fig:china} shows the min-max normalized volume of those hashtags over time; they peak during the first phase, when fear-related mentions and the focus of Twitter discussions on China and Asian foreigners were at their peaks.

\renewcommand{\thefigure}{S2}
\begin{figure}[h!]
	\centering
		\includegraphics[width=0.90\linewidth]{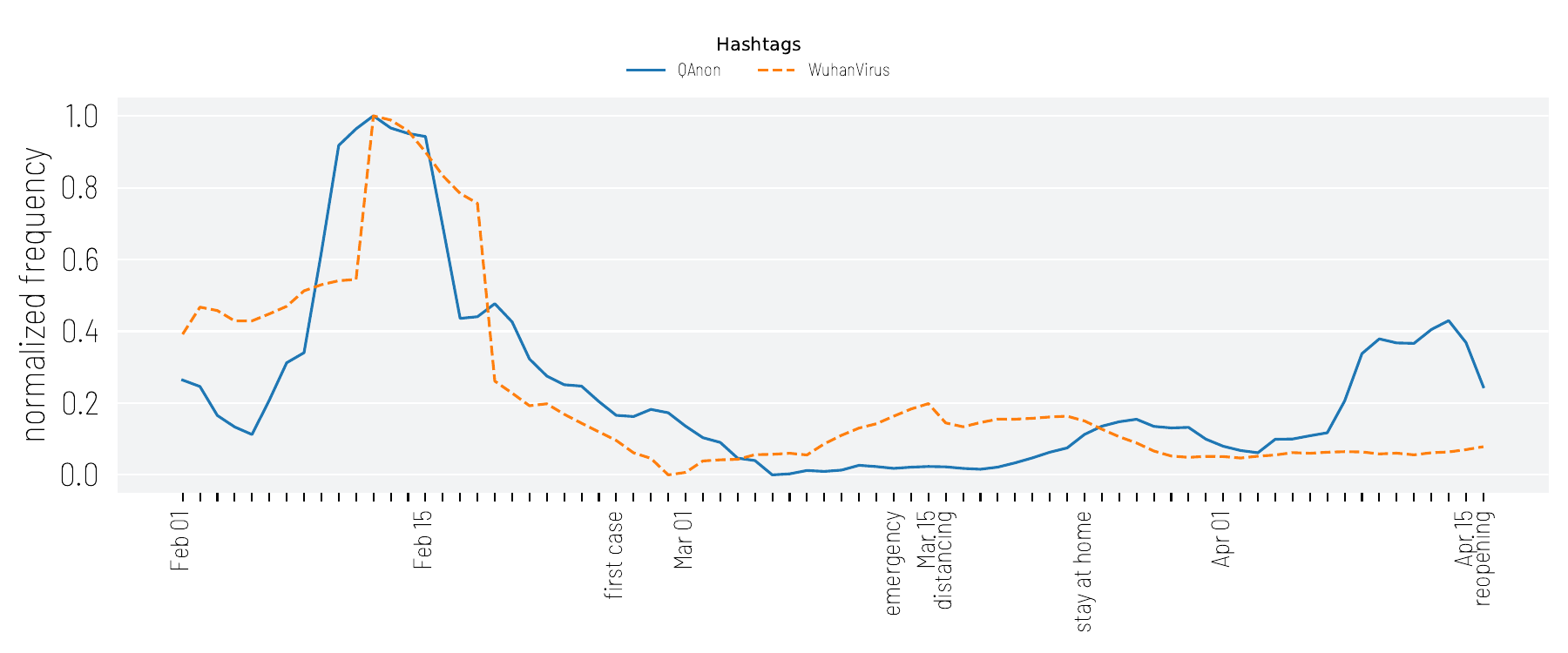}
	\caption{Min-max normalized frequency of hashtags connected to content that is xenofobic, racist, or generally aimed at antagonizing China during the COVID-19 pandemic. \#WuhanVirus characterizes tweets that claimed either that the virus was voluntarily caused by China or that it was an issue limited to China. \#QAnon denotes content supporting the so-called QAnon conspiracy that, during the first stages of the COVID pandemic, was amplifying narratives about Asian individuals being more vulnerable to the virus than white people, often using racist undertones.}
	\label{fig:china}
\end{figure}

%%%%%%%%%%%%%%%%%%%%%%%%%%%%%%%%%%%%%%%
%%%%%%%%%%%%%%%%%%%%%%%%%%%%%%%%%%%%%%%
\subsection*{Breakdown of phases by state}
%%%%%%%%%%%%%%%%%%%%%%%%%%%%%%%%%%%%%%%
%%%%%%%%%%%%%%%%%%%%%%%%%%%%%%%%%%%%%%%

During the period of analysis, the country was fragmented into regions that were dealing with different stages of the outbreak, and the policies to contrast it varied across states. To assess whether those differences impacted the unfolding of the psycho-social epidemics across states, we computed the measures that represent the relative presence of the three phases over time ($\Delta\%_{Refusal}$, $\Delta\%_{Anger}$, $\Delta\%_{Acceptance}$) for each state individually. The results are shown in Figure~\ref{fig:joyplot}. Despite minor differences, the state-level curves indicate that: \emph{i)} the three phases took place in all states; \emph{ii)} the sequence of the phases is consistent across states (refusal is followed by anger, which is then followed by acceptance); and \emph{iii)} there is very little variation of the times of transition between phases.

\renewcommand{\thefigure}{S3}
\begin{figure}[h!]
	\centering
		\includegraphics[width=0.99\linewidth]{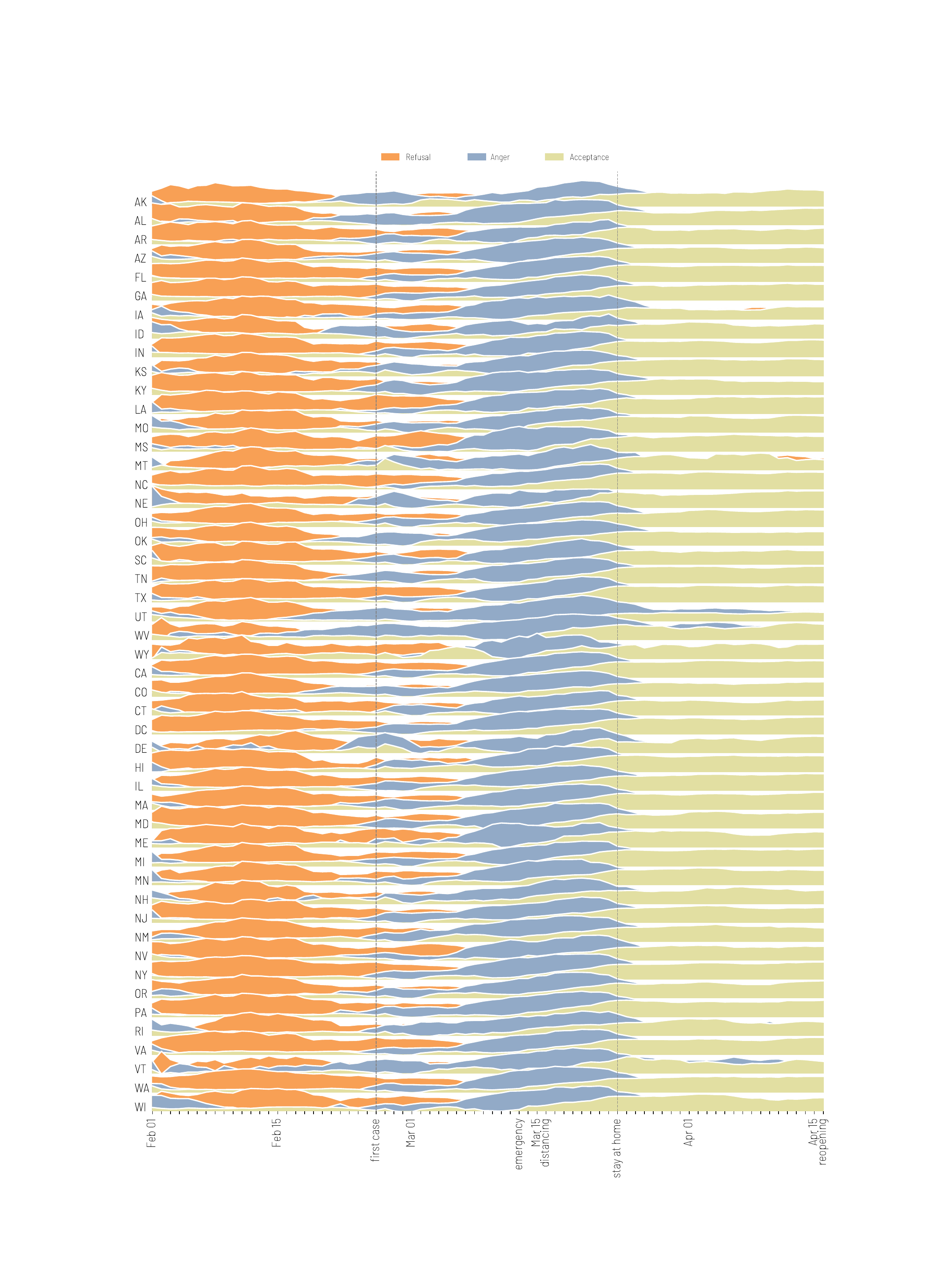}
	\caption{Breakdown by US states of the temporal evolution of  the relative frequency of a selected subset of the language categories associated with each of the three phases of refusal, anger, and acceptance.}
	\label{fig:joyplot}
\end{figure}

\end{document}